\documentclass[mathleft
]{an}
\usepackage{graphicx}
\usepackage{times}
\begin{document}


\title{Current-driven instabilities in weakly ionized disks}

\author{Edward Liverts\inst{1}\fnmsep\thanks{Corresponding author:
  \email{eliverts@bgu.ac.il}\newline}
Michael Mond\inst{1}
\and  Vadim Urpin\inst{2}
}
\titlerunning{Current-driven instabilities}
\authorrunning{E. Liverts, M. Mond \& V. Urpin}
\institute{
Department of Mechanical Engineering,  Ben-Gurion
University of the Negev, P.O. Box 653, Beer-Sheva 84105,
Israel
\and
A.F. Ioffe Institute of Physics and Technology, 194021 St. Petersburg, Russia}

\received{}
\accepted{}
\publonline{later}

\keywords{protoplanetary discs - accretion discs - MHD - Hall-MHD - instabilities}

\abstract{%
Cool weakly ionized gaseous rotating disk form the basis for many
models in astrophysics objects. Instabilities against
perturbations in such disks play an important role in the theory
of the formation of stars and planets. Traditionally, axisymmetric
magnetohydrodynamic (MHD) and recently Hall-MHD instabilities have
been thoroughly studied as providers of an efficient mechanism for
radial transfer of angular momentum, and of density radial
stratification. In the current work, the Hall instability against
axisymmetric  perturbations in incompressible rotating fluid in
external poloidal and toroidal magnetic field is considered.}

\maketitle

\section{Introduction}
The origin of turbulence in astrophysical disks is often attributed to
hydrodynamic and hydromagnetic instabilities that can occur in differentially
rotating stratified gas. The magnetorotational instability (MRI) is usually
considered as one of the possible candidates because it can operate in a
conductive flow if the angular velocity decreases with the cylindrical
radius (\cite{vel};~~\cite{chandra}).
The MRI has been studied in detail for both stellar
and accretion disk conditions (see, e.g.,
\cite{fricke};\cite{safronov};\cite{ach},1979;\cite{bh};
\cite{kai};\cite{zhang}
Note that the MRI occurs only in the presence of a weak magnetic
field, because a sufficiently strong field can suppress the instability
completely. Numerical simulations of the MRI in accretion disks (~\cite{haw};
~\cite{brand};~\cite{mats};
~\cite{tork};\cite{arlt})
show that the turbulence generated can enhance essentially the angular momentum transport.

The properties of turbulence in low conductive protostellar disks can
differ essentially from those of accretion disks. The magnetic Reynolds
number is not very large in protostellar disks and, hence, the field
cannot be treated as `frozen` into the gas (~\cite{gamm}). \,The influence of
ohmic dissipation on the MRI has been considered in the linear (\cite{jin})
and nonlinear regimes (~\cite{sano}). Calculations of many
authors indicate that the MRI is unlikely to be the source of turbulence in
protostellar disks since it arises only in a highly conductive plasma. For
instance, ~\cite{turn} have considered in detail turbulent mixing
caused by the MRI in protostellar disks and argued that the MRI does not
arise in the midplane even under the most favorable conditions because the
midplane is shielded from cosmic rays which are the main ionizing factor.
However, the number of instabilities that occur in disks is not
restricted by the MRI alone. An analysis of MHD modes in stratified disks
demonstrates a wide variety of instabilities even in the case of a very
simple magnetic geometry (~\cite{kepp}). More complex
magnetic geometries with a non-vanishing radial field lead to additional
instabilities (~\cite{bon}, 2008). Generally, even a pure
hydrodynamic origin of turbulence cannot be excluded (see, e.g., \cite{dubr};
\cite{lesur}).

It was first pointed out by \cite{wardle} that poorly conducting protostellar
disks can be strongly magnetized if electrons are the main charge carriers.
As a result, transport must be  anisotropic with substantially different
properties along and across the magnetic field. The effect of the magnetic
field on transport properties of plasma is characterized by the magnetization
parameter $a_{\rm e} = \omega_{B} \tau$ where $\omega_{B} = e B /m_{\rm e} c$
is the gyrofrequency of electrons and $\tau$ is their relaxation time (see,
e.g., \cite{spit}). In protostellar disks, $\tau$ is determined by the
scattering of electrons on neutrals and can be calculated by making use of
the fitting expression for the cross-section obtained by \cite{drai}.
Then, we have for the magnetization parameter
\begin{equation}
a_{\rm e} \approx 21\ B n_{14}^{-1}
T_{2}^{-1/2},
\label{magparam}
\end{equation}
where $B$ is the magnetic field measured in Gauss, $n_{14}= n/10^{14}$
cm$^{-3}$ and $T_{2} = T/100$ K with $n$ and $T$ being the number density of
neutrals and temperature, respectively. If $a_{\rm e} > 1$, i.e. $B > 0.048
\ n_{14} \sqrt{T_2}$ G, then the electron transport is anisotropic, and
the magnetic diffusivity is represented by a tensor. In a weakly ionized
plasma of protostellar disks, the difference between the
components of the magnetic diffusivity which are parallel and perpendicular
to the magnetic field is small (see, e.g., \cite{bt}), but the Hall
component that is perpendicular to the both magnetic field and electric
current can be much greater. The Hall component of diffusivity is given by
$a_{\rm e} \eta$ where $\eta=c^{2} m_{\rm e}/ 4 \pi e^{2} n_{\rm e} \tau$
is the magnetic diffusivity
at $B=0$; $m_{\rm e}$ and $n_{\rm e}$ are the mass and number density of
electrons, respectively. Using the fit for the cross-section by \cite{drai},
we obtain for the magnetic diffusivity
\begin{equation}
\eta = 2.34 \times 10^{3} x_e^{-1} T_{2}^{1/2} \;\; {\rm cm}^{2} \;
{\rm s}^{-1},
\end{equation}
where $x_e=n_{\rm e}/n$ is the ionization fraction.

A stability analysis of the MRI done by \cite{wardle} shows that the Hall
effect can provide either stabilizing or destabilizing influence depending on
the direction of the field. A more general consideration of the Hall MRI has
been done by Balbus \& Terquem (2001). They found that the Hall effect changes
qualitatively the stability properties of protostellar disks and can lead to
instability even if the angular velocity increases outward. These authors,
however, did not take into account the effect of gravity that is crucial for
disks. A consistent consideration of the linear MRI under the combined
influence of the Hall effect and gravity has been done by \cite{urp}
who derived also the criteria of several other instabilities that can
occur in protostellar disks. The properties of the MRI modified by the Hall
effect has been considered also by \cite{salm}. These authors
argued that the MRI is active in protoplanetary disks over a wide range of
field strengths and fluid conditions. The Hall conductivity results in a
faster growth of perturbations and extends the region of instability. Recently,
\cite{liv} and \cite{shtem} have considered the Hall instability (HI)
for non-axisymmetric perturbations. This instability differs
from the MRI and it results from the fast magnetosonic waves in contrast
to the Alfven nature of the MRI. The HI instability is proposed as a
viable mechanism for the azimuthal fragmentation of the protoplanetary disks
and planet formation. The non-axisymmetric instability is caused basically by
the combined effect of the radial stratification and Hall electric field.

Apart from the Hall effect, the stability properties of magnetic protostellar
disks can be influenced by a number of other factors, for example, the
electric currents. In differentially rotating disks, the azimuthal field is
generated by stretching from the poloidal one, and poloidal currents
are necessary to maintain this azimuthal field that typically is not
current-free. The generated azimuthal field can be stronger than the
poloidal one if the magnetic Reynolds number is larger than 1. The azimuthal
field and associated currents can be important for stability of disks even
if the Hall effect is negligible (see \cite{pess}). In
protostellar disks, the effect of currents maintaining the magnetic
configuration is accompanied often by the Hall effect that changes crucially
the stability properties.

~In the present paper we consider the linear stability properties of magnetic
protostellar disks taking into account the combined influence of the Hall
effect and electric currents. The criteria of instability are derived, and
the growth rate of the various modes is calculated.

\section{Basic equations}

Consider the stability of a magnetized protostellar disk of a finite vertical
extent. For the sake of simplicity, the unperturbed angular velocity is
assumed to be dependent on the cylindrical radius $s$ alone such as $\Omega
= \Omega(s)$; ($s$, $\varphi$, $z$) are cylindrical coordinates. The magnetic
field, $\vec{B}=(B_{s}, B_{\varphi}, B_{z})$, is assumed to be weak in the sense
that the Alfv\'en speed, $c_{\rm A}$, is small compared to the sound speed,
$c_{\rm s}$. This enables us to employ the Boussinesq approximation for a
consideration of slowly varying modes. In the unperturbed state, the disk
is assumed to be in hydrostatic equilibrium in the $s$- and $z$-directions,
\begin{equation}
\frac{\nabla p}{\rho} = \vec{G} + \frac{1}{4 \pi \rho}
{\rm rot} \vec{B} \times \vec{B} \;\;, \;\;\;\;
\vec{G} = \vec{g} + \Omega^{2} \vec{s} \;,
\label{hydrostatic}
\end{equation}
here $\vec{g}$ is the gravity force per unit mass. It should be noticed
however that the pressure gradient in the unperturbed disks is mainly determined
by gravity and centrifugal forces. The unperturbed Lorentz force
namely the second term on the r.h.s. Eq.~(\ref{hydrostatic}) is usually much
smaller than gravity and centrifugal forces (the first term on the r.h.s. of the
last equation) and therefore may be neglected. This is due to the assumption
of small thickness of the disk. Detailed asymptotic description of that fact
may be found in \cite{regev};\cite{klu};\cite{shtem}.

We consider the stability of axisymmetric short wavelength perturbations with
space-time dependence $\exp(\gamma t - i \vec{k} \cdot \vec{x})$ where
$\vec{k}= (k_{s}, 0, k_{z})$ is the wave vector. ~The linearized momentum and
continuity equations read in the Boussinesq approximation
{\setlength{\mathindent}{0pt}
\begin{eqnarray}
\gamma \vec{v} + 2 \vec{\Omega} \times \vec{v} +
\vec{e}_{\varphi} s (\vec{v} \cdot \nabla) \Omega =
\frac{i \vec{k} p}{\rho} - \alpha \vec{G} T_{1} +
\nonumber \\
\frac{i}{4 \pi \rho} [(\vec{B} \cdot \vec{b})\vec{k} - (\vec{k}
\cdot \vec{B})\vec{b}]-\frac{\vec{e}_r}{4 \pi \rho}\frac{B_\varphi
b_{\varphi}}{s} + \frac{1}{c \rho} \vec{J} \times
\vec{b}, \label{momentumeq}\\
\vec{k} \cdot \vec{v} = 0 , \label{incompres}
\end{eqnarray}}
where $\vec{v}$, $\vec{b}$, $p$ and $T_{1}$ are the perturbations
of the hydrodynamic velocity, magnetic field, pressure and temperature,
respectively; $\alpha= - (\partial \ln \rho / \partial T)_{P}$ is the thermal
expansion coefficient and $\vec{e}_{\varphi}$ is the unit vector in the azimuthal
direction. It is assumed in Eq.~(\ref{momentumeq}) that the density perturbation in the
buoyancy force is determined by the temperature perturbation alone in
accordance with the main idea of the Boussinesq approximation, $\rho_{1}= -
\rho \alpha T_{1}$, and the unperturbed Lorentz force in
Eq.~(\ref{hydrostatic})is neglected. We took into account the effect of electric currents (the
last term on the r.h.s. of Eq.~(\ref{momentumeq})). In short wavelength approximation, this
term seems to be smaller than the previous one by a factor $\sim kL \gg 1$,
where $L$ ($\sim s$) is the length-scale of the unperturbed magnetic field.
In differentially rotating discs, however, the electric current $\vec{J}=
(c/4 \pi) \nabla \times \vec{B}$ is mainly determined by the $\varphi$-component
of the magnetic field that can be substantially stronger than $B_{s}$ and
$B_{z}$ because of stretching the magnetic field lines in the azimuthal
direction by differential rotation. If $B_{\varphi}$ satisfies the condition
\begin{equation}
B_{\varphi} > kL \max(B_{s}, B_{z}),
\label{cond}
\end{equation}
then the effect of electric currents cannot be neglected in Eq.~(\ref{momentumeq}). ~For the
sake of simplicity, we assume that the toroidal field $B_{\varphi}$ depends on
$s$ alone, then
\begin{equation}
\vec{J} = J_{z} \vec{e}_{z} \;, \;\; J_{z} =
\frac{c}{4 \pi s} \frac{\partial}{\partial s} (s B_{\varphi}).
\end{equation}
Note that, calculating the perturbation of the electric current $\vec{j} =
(c/4 \pi ) \nabla \times \vec{b}$ in Eq.~(\ref{momentumeq}), we can neglect terms of
the order of $1/s$ and assume $\vec{j} \approx  - (ic/4 \pi) \vec{k}
\times \vec{b}$. Using Eq.~(\ref{incompres}), we can calculate $p$ from Eq.~(\ref{momentumeq}).
Then, we obtain for the momentum equation
\begin{eqnarray}
\gamma \vec{v} + 2 \vec{\Omega} \times \vec{v}
- \frac{2 \vec{k}}{k^{2}} \vec{k} \cdot (\vec{\Omega} \times \vec{v})
+ \vec{e}_{\varphi} s \Omega' v_{s} = \nonumber \\
- \alpha T_{1} \left[ \vec{G} - \frac{\vec{k}}{k^2} (\vec{k} \cdot
\vec{G}) \right]
- \frac{i}{4 \pi \rho} (\vec{k} \cdot \vec{B})\vec{b}\nonumber \\
+ \frac{1}{c \rho} \left[ \vec{J} \times \vec{b} - \frac{\vec{k}}{k^2}
\vec{k} \cdot (\vec{J} \times \vec{b}) \right].
\label{momentumeq1}
\end{eqnarray}
It is clearly seen from the last equation that the third term on the r.h.s.
which results from the unperturbed electric current can be larger under condition (\ref{cond})
than the second term which is usually taken into account in a stability
analysis of disks and which contains only poloidal components $B_{s}$
and $B_{z}$ of the background field.

Since the thermal conductivity of protostellar disks is low because
of a low temperature ($T \sim 10-10^{3}$ K), we adopt the adiabatic
equation to describe the evolution of temperature perturbations,
\begin{equation}
\gamma T_{1} + \vec{v}\cdot (\Delta \nabla T) = 0 \;,
\label{7}
\end{equation}
where $(\Delta \nabla T) = \nabla T - \nabla_{\rm ad} T $ is the
difference between the actual and adiabatic temperature gradients.
Substituting $T_{1}$ into Eq.~(\ref{momentumeq1}), we obtain the equation that
contains only perturbations of $\vec{v}$ and $\vec{b}$.
The $s$- and $\varphi$-components of this equation read
{\setlength{\mathindent}{0pt}
\begin{eqnarray}
(\gamma + \frac{\omega_{g}^{2}}{\gamma}) v_{s} -
2 \mu \Omega v_{\varphi} =
- \frac{i}{4 \pi \rho}(\vec{k}
\cdot \vec{B}) b_{s} - \frac{\mu J_{z}}{c \rho}
b_{\varphi}, \label{momentums} \\
\gamma v_{\varphi} + (2 \Omega + s \Omega') v_{s} =
- \frac{i}{4 \pi \rho} (\vec{k} \cdot \vec{B}) b_{\varphi} +
\frac{J_{z}}{c \rho} b_{s}, \label{momentumphi}
\end{eqnarray}}
where
$$
\omega_{g}^{2} = - \alpha \Delta \nabla T \cdot \left[ \vec{G} -
\frac{\vec{k}}{k^{2}} (\vec{k} \cdot \vec{G}) \right]
$$
and $\mu = k_{z}^{2}/k^{2}$.

As it was mentioned, the effect of the magnetic field on kinetic properties
of plasma is usually characterized by the magnetization parameter $a_{\rm e}$
(see Eq.~(\ref{magparam})). We consider the most interesting case for protoplanetary disks
when this parameter is moderate, $kL \gg a_{e}$. Under this assumption, the
linearized induction equation reads
{\setlength{\mathindent}{0pt}
\begin{eqnarray}
\gamma \vec{b} &=&-\eta \nabla \times (\nabla \times \vec{b})
+ \nabla \times (\vec{v} \times \vec{B})
+ \nabla \times (s \Omega \vec{e}_{\varphi} \times \vec{b})
\nonumber \\
&-&\frac{c}{4 \pi e n_{e}} \nabla \times [ (\nabla \times \vec{b})
\times \vec{B} + (\nabla \times \vec{B}) \times \vec{b}],
\label{inductioneq}
\end{eqnarray}}
~~where $\eta$ is magnetic diffusivity and $n_{e}$ is the number density
of electrons. ~~For short wavelength perturbations, ~~~~Eq.~(\ref{inductioneq})
and the divergence-free condition read
{\setlength{\mathindent}{0pt}
\begin{eqnarray}
(\gamma + i \omega_{\wedge}) \vec{b} = -i \vec{v}(\vec{k}
\cdot \vec{B}) - \vec{e}_{\varphi} v_{s} s \frac{\partial}{\partial s}
\left(\frac{B_{\varphi}}{s}\right) + \vec{e}_{\varphi} s \Omega' b_{s}
\nonumber\\
-\vec{e}_{z}\frac{J'_{z}}{e n_{e}} b_{s} + \frac{c(\vec{k}\cdot \vec{B})}{4 \pi e n_{e}}
\vec{k} \times \vec{b} +
\vec{e}_{\varphi}\frac{i c k_{z}}{4 \pi e n_{e}} B'_{\varphi} b_{\varphi},
\label{inductioneq1}\;\;\;\\
\vec{k}\cdot\vec{b} = 0, \;\;\;\;\label{free}
\end{eqnarray}}
where $\omega_{\wedge} = k_{z} J_{z}/ e n_{e} - i \eta k^{2}$,
$B'_{\varphi}=d B_{\varphi}/ds$, and $J'_{z} = d J_{z}/d s$; the last
three terms on the r.h.s. of Eq.~(\ref{inductioneq1}) together with the first term in
$\omega_{\wedge}$ represent the Hall effect that can be important in
protoplanetary disks. Under condition (\ref{cond}), induction equation
(\ref{inductioneq1}) can be simplified because the fourth term on the r.h.s is small
compared to the last term and do not influence the behavior of perturbations.
Then, we have for the induction equation
{\setlength{\mathindent}{0pt}
\begin{eqnarray}
{\setlength{\mathindent}{0pt}
(\gamma + i \omega_{\wedge}) \vec{b} =
- i \vec{v} (\vec{k} \cdot \vec{B})-\nonumber}\\
\vec{e}_{\varphi}  \left[ v_{s} s \frac{\partial}{\partial s}
\left( \frac{B_{\varphi}}{s} \right) - s \Omega' b_{s}  -
i \omega_{H} b_{\varphi} \right] + \frac{c
(\vec{k} \cdot \vec{B})}{4 \pi e n_e} \vec{k}\times\vec{b},\;\;\;\;
\end{eqnarray}}
where $\omega_{H} = c k_{z} B'_{\varphi}/4 \pi e n_{e}$. The $s$- and
$\varphi$-components of this equation are
{\setlength{\mathindent}{0pt}
\begin{eqnarray}
(\gamma + i \omega_{\wedge}) b_{s} = - i (\vec{k} \cdot \vec{B})
v_{s} +
\omega_{w} b_{\varphi}, \label{inductions}\\
(\gamma + i \omega_{0}) b_{\varphi} = - i (\vec{k} \cdot \vec{B})
v_{\varphi} - s \frac{\partial}{\partial s} \left(
\frac{B_{\varphi}}{s} \right) v_{s} - \nonumber \\
\frac{i \vec{k} \cdot
\vec{B}}{\gamma + i \omega_{\wedge}} s \Omega' v_{s} -
\frac{k^2}{k_z^2} \omega_{w} b_{s} \;,\label{inductionphi}
\end{eqnarray}}
where
{\setlength{\mathindent}{0pt}
\begin{equation}
\omega_{0} = \omega_{\wedge} - \omega_{H}=\frac{c k_z B_{\varphi}}{2
\pi e n_e s} - i \eta k^2, \omega_{w}= \frac{c k_z (\vec{k}
\cdot \vec{B})}{4 \pi e n_e}.
\end{equation}}
Eqs.~(\ref{momentums}), (\ref{momentumphi}), (\ref{inductions}),
and (\ref{inductionphi}) describe the eigenmodes that exist in a
magnetized fluid in the presence of electric currents.

\section{Dispersion equation and stability criteria}

The general dispersion equation for the set of
Eqs.~(\ref{momentums}), (\ref{momentumphi}), (\ref{inductions}),
and (\ref{inductionphi})
is rather cumbersome. Therefore, we consider only a particular case of
perturbations with the wavevector perpendicular to the unperturbed magnetic
field, $\vec{k} \cdot \vec{B} = 0$. The MRI does not occur for such
perturbations, and they should be stable if the Hall effect is neglected and
$\vec{J} = 0$. Therefore, the  particular case $\vec{k} \cdot \vec{B} = 0$
allows to study a destabilizing influence of both the unperturbed electric
current and the Hall effect in a situation when other factors can cause
only a stabilizing influence. If $\vec{k} \cdot \vec{B} = 0$, then the
dispersion equation takes the form
\begin{equation}
\gamma^3 + a_2 \gamma^2 + a_1 \gamma + a_0 = 0,
\label{disp}
\end{equation}
where
{\setlength{\mathindent}{0pt}
\begin{equation}
a_2 = i \omega_0, a_1 = \omega_g^2 + \mu (\kappa^2
+ \omega_J^2), a_0 = i \omega_0 ( \omega_g^2
+ \mu \kappa^2).
\end{equation}}
The characteristic frequencies in these expressions are given by
\begin{eqnarray}
\omega_0 = \omega_1 - i \omega_{\eta}, \;\; \omega_1 =
\frac{c k_z B_{\varphi}}{2 \pi e n_e s}, \;\;
\omega_{\eta} =\eta k^2, \;\; \nonumber\\
\kappa^2 = 2 \Omega (2 \Omega + s \Omega'), \;\;
\omega_{J}^2 = -\frac{s J_z}{c\rho} \frac{\partial}{\partial s} \left(
\frac{B_{\varphi}}{s} \right)
\end{eqnarray}
For $\vec{k} \cdot \vec{B}=0$, only three non-trivial modes exist in the flow but
other modes are degenerate.

\subsection{Stability in the case $a_e \ll 1$}

If the Hall parameter is small, the Hall effect does not influence
the stability properties. In this case, $\omega_0 \approx - i \omega_{\eta}$.
Then, the dispersion equation takes the form
\begin{equation}
\gamma^3 + b_2 \gamma^2 + b_1 \gamma + b_0 = 0,
\label{disp1}
\end{equation}
where all coefficients of this equation are real,
\begin{eqnarray}
b_2 = \omega_{\eta}, \;\;\; b_1 = a_1 = \omega_g^2 +\mu (\kappa^2
+ \omega_J^2), \;\;\; \nonumber\\ b_0 = \omega_{\eta} ( \omega_g^2 +
\mu \kappa^2 ).
\label{23}
\end{eqnarray}
The condition that at least one of the roots of Eq.~(\ref{disp1}) has a positive
real part (that corresponds to instability) is equivalent to one of the
following inequalities
\begin{equation}
b_2 < 0, \;\;\; b_1 b_2 < b_0, \;\;\; b_0 < 0
\end{equation}
being fulfilled (see, e.g., \cite{aleks}). Since $\omega_{\eta}
> 0$, the first condition $b_2 < 0$ will never apply. The other two
conditions yield
\begin{eqnarray}
\mu \omega_{\eta} \omega_{J}^2 < 0  \;\; , \;\;\;
\omega_{\eta} (\omega_g^2 + \mu \kappa^2) < 0.
\label{25}
\end{eqnarray}
Both criteria are proportional to the dissipative frequency (that is positively
defined quantity) and appear only if one takes into account magnetic
diffusivity. Therefore, both criteria describe diffusive instabilities that
can be relatively fast in protostellar disks. In ideal magnetohydrodynamics,
we have $\omega_{\eta} =0$ and $b_0 = b_2= 0$, and dispersion relation (\ref{disp})
transforms into
\begin{equation}
\gamma^2 + a_1 =0.
\end{equation}
The condition of instability reads $a_1 <0$, or
\begin{equation}
\omega_g^2 + \mu (\kappa^2 + \omega_J^2) < 0,
\label{27}
\end{equation}
that differ substantially from dissipative criteria (\ref{25}). For instance,
to satisfy condition (\ref{27}) in disks, energy of the magnetic field should be
comparable to the rotational or gravitational energy of gas. The first
condition (\ref{25}) can be fulfilled even in a much weaker field. Since
$\omega_{\eta} >0$ and $\mu>0$, we obtain from Eq.~(\ref{25}) the following
conditions of instability
\begin{eqnarray}
\omega_{J}^2 < 0 , \label{28}\\
\omega_g^2 + \mu \kappa^2 < 0.
\label{29}
\end{eqnarray}
Eq.~(\ref{29}) is the standard criterion of convection modified by rotation and
usually is not satisfied in astrophysical discs. Eq.~(\ref{28}) is the condition
of an instability that can occur due to the presence of electric currents
(see also \cite{vel}).
This instability is associated only with the distribution of electric
currents in disks. Condition (\ref{28}) can be rewritten as
\begin{equation}
B_{\varphi}^2 - s^2 B_{\varphi}'^2 < 0.
\label{30}
\end{equation}
Therefore, the current-driven instability arises if ~$B_{\varphi}$ decreases
with $s$ faster than $1/s$ or increases outward faster than $s$. Note also
that the current-driven instability does not depend on the angular velocity
profile and can occur for perturbations that are not subject to the MRI
since $\vec{k} \perp \vec{B}$.

Since the coefficients of equation (\ref{disp1}) are real there exist three real roots
or one real and two complex conjugate roots. The number of roots with a
positive real part is determined by Routh criterium (DiStefano III,
Stubberud \& Williams 1994), which states that the number of unstable modes
of a cubic equation (\ref{disp1}) is given by the number of changes of sign in the
sequence
\begin{equation}
\left\{ 1, \; b_{2}, \; \frac{b_{2} b_{1} - b_{0}}{b_{2}},
\; b_{0} \right\}.
\end{equation}
For coefficients (23), this sequence reads
\begin{equation}
\{1 \;, \;\; \omega_{\eta} \;, \;\; \omega_{J}^{2} \;, \;\;
\omega_{\eta} (\omega_g^{2} + \mu \kappa^2) \}.
\end{equation}
If the disc is convectively stable ($\omega_{g}^{2} + \mu \kappa^2 > 0$), we
obtain that under the condition (\ref{28}) (or (\ref{30})) two complex conjugate modes
are unstable. If the disc is convectively unstable ($\omega_{g}^{2} + \mu
\kappa^2 < 0$) and the condition (\ref{28}) holds then there should be only one
unstable mode. Only one mode is unstable also in the case when the disc is
convectively unstable but the condition (\ref{28}) is not fulfilled.

The roots $\gamma_{i}$ ($i=1,2,3$) of the cubic equation (\ref{disp}) can be
represented as $\gamma_{i} = x_{i} - a_{2}/3$. The expressions for $x_{i}$
are
\begin{equation}
x_{1} = u+v \;, \;\; x_{2, 3}= - \frac{1}{2} (u + v) \pm i \frac{\sqrt{3}}{2}
(u -v ) \;,
\label{33}
\end{equation}
where
\begin{eqnarray}
(u, v) = (-q \pm \sqrt{q^{2}+p^{3}})^{1/3} \;, \nonumber\\
2 q = \frac{2}{27} a_2^3 - \frac{1}{3} a_2 a_1 + a_0 \; , \;\;
3 p = a_1 - \frac{1}{3} a_2^2 \;.
\label{34}
\end{eqnarray}
(see, e.g., \cite{BS}). In a particular case $a_{\rm e}<1$
when the coefficients of a cubic equation are given by Eq.~(\ref{23}), we have
\begin{eqnarray*}
q = \frac{\omega_{\eta}}{3} \left( \omega_{g}^{2} + \mu \kappa^2
-\frac{1}{2} \mu \omega_{J}^{2} + \frac{1}{9} \omega_{\eta}^2
\right), \;\;\;\\
p = \frac{1}{3} \left( \omega_{g}^{2} + \mu \kappa^2 + \mu \omega_{J}^2
- \frac{1}{3} \omega_{\eta}^2
\right).\;\;\;\;
\end{eqnarray*}

In the limit of small $\omega_{\eta}$, we have for the roots
\begin{equation}
\gamma_{1} = - \frac{\omega_{\eta}(\omega_g^2 + \mu \kappa^2)}{\omega_{g}^{2}
+ \mu \kappa^2 + \mu \omega_{J}^2} ,
\label{35}
\end{equation}
{\setlength{\mathindent}{0pt}
\begin{equation}
\gamma_{2, 3} = \pm i \sqrt{ \omega_{g}^{2} + \mu \kappa^2 + \mu \omega_{J}^2}
- \frac{1}{2} \frac{\omega_{\eta} \mu \omega_{J}^2}{\omega_{g}^{2} +
\mu \kappa^2 + \mu \omega_{J}^2}.
\label{36}
\end{equation}}
In this case, the instability is dissipative since the growth rate
is proportional to $\eta$.

If $\omega_g^2 + \mu (\kappa^2 + \omega_J^2) > 0$, then there should be no
instability in the ideal magnetohydrodynamics (see condition (\ref{27})). Indeed,
expressions (\ref{35}) and (\ref{36}) yield $\rm{Re} \gamma =0$ in the limit
$\omega_{\eta} \rightarrow 0$. However, in a dissipative MHD, the instability
can occur even if $\omega_g^2 + \mu (\kappa^2 + \omega_J^2) > 0$.
In this case, the first mode (non-oscillatory) is unstable alone if
$\omega_g^2 + \mu \kappa^2 < 0$, but two oscillatory modes are stable since
$\omega_J^2$ should be positive. On the contrary, if $\omega_g^2 +
\mu (\kappa^2 + \omega_J^2) > 0$ but $\omega_J^2 < 0$, then two oscillatory
modes are unstable, but the first mode should be stable since $\omega_g^2 +
\mu \kappa^2$ is positive. If $\omega_g^2 + \mu (\kappa^2 + \omega_J^2) <
0$, then mode 3 is unstable with a very large growth rate $\gamma_3 =
\sqrt{|\omega_g^2 + \mu (\kappa^2 + \omega_J^2)|}$ but mode 2 is
rapidly decaying with the decay rate approximately equal to $\gamma_3$. The
first mode can also be unstable in this case if $\omega_g^2 + \mu \kappa^2 >
0$ but its growth rate is small since it is proportional to the small
dissipative frequency $\omega_{\eta}$.

If $\omega_{\eta}$ is greater than other characteristic frequencies, then
we have from Eqs.~(\ref{33})-(\ref{34})
\begin{equation}
\gamma_{1} \approx -\omega_{\eta}, \quad \gamma_{2, 3} \approx \pm i
\sqrt{\omega_{g}^{2} + \mu \kappa^2} -
\frac{\mu \omega_{J}^2}{2\omega_{\eta}}.
\label{37}
\end{equation}
Mode 1 is always stable but modes 2 and 3 can be unstable. If
$\omega_g^2 + \mu \kappa^2 > 0$, then modes 2 and 3 are oscillatory.
The instability of these modes occurs if $\omega_{J}^2 < 0$. In the opposite
case $\omega_J^2 > 0$, oscillatory modes do not arise. If $\omega_g^2 +
\mu \kappa^2 < 0$, then oscillatory modes become non-oscillatory, and one
of these modes is unstable.
The case of large $\omega_{\eta}$ corresponds to small magnetic Reynolds
number and is of particular interest for protostellar disks. The instability
of oscillatory modes can operate even in the dead zone of protoplanetary
disks where the magnetic Reynolds number is small.

\subsection{Stability in a strong magnetic field with $a_e \geq 1$}

Let us consider the stability of a strongly magnetized plasma with ~~$a_e
\geq 1$. In this case, the growth rate is described by Eq.~(\ref{disp}) with complex
coefficients. The roots of Eq.~(\ref{disp}) can be calculated by making use of general
expressions \\ (\ref{33})-(\ref{34}) for the roots of a cubic equation. \,\,\,However, these
expressions are rather cumbersome and \,\,inconvenient for analysis. Therefore,
we consider in detail the growth rate in the case when the frequency
associated to electric currents $\omega_J$ is lower than the angular
velocity $\Omega$ or characteristic buoyancy frequency $\omega_g$. The
dissipative frequency $\omega_{\eta}$ can be high and comparable to (or even
higher than) other characteristic frequencies. This case is of particular
interest for the dead zones of protostellar disks where the conductivity is
extremely low, and the magnetic Reynolds number can be relatively small.
We can rewrite Eq.~(\ref{disp}) as
\begin{equation}
(\gamma^2 + \omega_g^2 + \mu \kappa^2) + \frac{\gamma \mu \omega_{J}^2}{\gamma
+ i \omega_0} = 0.
\label{disp3}
\end{equation}
This shape is more convenient to calculate the oscillatory roots (which can
generally be unstable) by making use of a perturbation procedure. Since the
last term on the l.h.s. is proportional to the square of a low frequency
$\omega_J$, it can be considered as a small perturbation. Therefore, the
solution of Eq.~(\ref{disp3}) can be represented as a power series of $\omega_J^2$:
$\gamma = \gamma^{(0)} + \gamma^{(1)} + ...$ where $\gamma^{(0)}$ does not
depend on $\omega_{J}^2$ and $\gamma^{(1)}$ is linear in $\omega_{J}^2$. The
equation of the zeroth order yields
\begin{equation}
\gamma^{(0)} = \pm i \sqrt{\omega_g^2 + \mu \kappa^2}
\end{equation}
(we assume that the unperturbed disk is convectively stable and $\omega_g^2
+ \mu \kappa^2 > 0$). The roots are imaginary in the zeroth approximation,
and there is no instability in the absence of electric currents. The
correction of the first order is
\begin{equation}
\gamma^{(1)} = - \frac{1}{2} \frac{\mu \omega_J^2}{\gamma^{(0)} + i \omega_0}.
\label{40}
\end{equation}
Splitting this equation into real and imaginary parts, we have for the
growth rate
\begin{equation}
{\rm Re} \gamma = - \frac{1}{2} \frac{\omega_{\eta} \mu \omega_J^2}{(\omega_1
\pm \sqrt{\omega_g^2 + \mu \kappa^2})^2 + \omega_{\eta}^2}.
\end{equation}
Like the case of a weakly magnetized disk, the instability arises only if
the magnetic field satisfies condition (\ref{28}). The growth rate depends on the
wavelength of perturbations and can be essentially different for different
$k$. If the wavelength $\lambda = 2 \pi / k$ is sufficiently short such as
$\omega_{\eta} > \Omega$, then the growth rate is approximately given by
\begin{equation}
{\rm Re} \gamma \approx - \frac{\mu \omega_J^2}{\omega_{\eta}}.
\label{42}
\end{equation}
One should note that the last expression (obtained for
a case of strongly magnetized electrons) is similar to the
one found in section 3.1 [see Eq.(\ref{37})]. It is assumed here that a dissipative
frequency $\omega_{\eta}$ is greater than all other characteristic frequencies.
The order of magnitude estimate of Re$\gamma$ is
\begin{equation}
{\rm Re} \gamma \sim 4.3 \times 10^{-5} \frac{B_{\varphi 2}^2 {x_e}_{-12}}{n_{14}
T_{2}^{1/2}} \left( \frac{\lambda}{s} \right)^2 \;\; {\rm s}^{-1},
\end{equation}
where $B_{\varphi 2} = B_{\varphi}/ 100$G and ${x_e}_{-12} = x_e/10^{-12}$.
The condition $\omega_{\eta} > \Omega$ is equivalent to
\begin{equation}
\lambda_{12} > 0.66 T_{2}^{1/2} P_{yr} {x_e}^{-1}_{-12},
\end{equation}
where $P_{yr}$ is the rotation period in years and $\lambda_{12} = \lambda/
10^{12}$ cm.

The imaginary part of $\gamma$ that determines the frequency of perturbations
is approximately given by expression of the zeroth order (\ref{40}). This
expression describes buoyancy waves modified by differential rotation. It
is well known that the buoyancy waves can be unstable and are responsible
for convection if $\omega_g^2 < 0$ that is not likely to be satisfied in
protostellar disks. The instability considered in this section is the
instability of buoyant waves as well and, in fact, is an oscillatory
modification of convection. In stellar hydrodynamics, an oscillatory
convection is often called semiconvection and can be caused, for example, by
a gradient of the chemical composition, As it is seen from our consideration,
the distribution of the toroidal field can also be the reason of
semiconvection in protostellar disks.

In the opposite case, $\omega_{\eta} << \Omega$, we have from Eq.~(\ref{42}) for
the growth rate
\begin{equation}
{\rm Re} \gamma \approx - \frac{1}{2} \; \frac{\omega_{\eta} \mu
\omega_J^2}{\omega_g^2 + \mu \kappa^2}.
\end{equation}
The instability can occur for perturbations with $\omega_{\eta} < \Omega$
as well, however, the growth rate is low in this case. The frequency of
such weakly unstable perturbations is given by Eq.~(\ref{40}).

\section{Summary and discussion}

This paper examines the instability of weakly ionized protostellar disks
threaded by an external magnetic field. The conductivity of such disks is
low, and dissipative effects can play an important role in the evolution
of perturbations. This concerns particularly the midplane that is shielded
from cosmic rays to such extent that the MRI does not occur even under the
most favorable conditions (Turner et al. 2006). Since the problem of
stability is rather cumbersome with taking account of dissipative effects,
we have considered a special case of perturbations with the wavevector
perpendicular to the magnetic field. Such perturbations are not subject to
the magnetorotational instability, because its growth rate is proportional
to $(\vec{k} \cdot \vec{B})$ and is vanishing for the considered
perturbations.

It turns out that dissipative process can alter drastically the stability
properties of protostellar disks. Apart from the instabilities that are
typical for non-dissipative differentially rotating disks, new instabilities
can occur that are determined by dissipative processes. In the simplest case
considered in this paper, condition of the dissipative instability (\ref{30}) is
entirely determined by the radial dependence of the azimuthal field and
can be represented as
\begin{equation}
\left| \frac{d \ln B_{\varphi}}{d \ln s} \right| > 1.
\end{equation}
The instability arises if the magnetic field decreases with $s$ faster than
$1/s$ or increases more rapidly than $s$. Note that the condition of the
considered instability does not depend on the magnetic diffusivity, and can
be satisfied in the limiting cases of very high ($\eta \rightarrow 0$) and
very low conductivity ($\eta \rightarrow \infty$). However, the growth rate
of instability depends sensitively on conductivity and is proportional to
$\eta$ and $1/\eta$ in high- and low-conductivity limits, respectively.

The condition of instability (\ref{37}) does not depend directly on the rotation
law. However, differential rotation can influence this condition indirectly
because the radial profile of $B_{\varphi}$ depends on $\Omega(s)$. If
stretching of the azimuthal field lines in the basic state is balanced
by ohmic dissipation, then we approximately have from the induction equation
\begin{equation}
\eta \Delta B_{\varphi} \sim s \Omega' B_s,
\end{equation}
or
\begin{equation}
B_{\varphi} \sim \frac{s^3 \Omega' B_s}{\eta}.
\end{equation}
If rotation is Keplerian and $\Omega \propto s^{-3/2}$, then the radial
dependence of $B_{\varphi}$ is given by $B_{\varphi} \propto s^{1/2} B_s /
\eta$. Therefore, the considered instability may occur in the disk if
the ratio $B_s / \eta$ decreases with $s$ faster than $s^{-3/2}$. Likely,
this condition can often be fulfilled in protostellar disks, particularly
if the radial field component decreases as a dipole ($\propto s^{-3}$).
Therefore, the considered instability can generate turbulence in regions
with a very low conductivity including the dead zones which likely exist in
protostellar disks.





\end{document}